\documentclass[twocolumn,showpacs,prb,amsmath,amssymb]{revtex4}
\usepackage{graphicx}
\usepackage{dcolumn}
\usepackage{bm}

\begin{document}
\title{Magnetic Inhomogeneity and Magnetotransport in Electron-Doped Ca$_{1-x}$La$_x$MnO$_3$
($0\leq x\leq 0.10$)}
\author{C. Chiorescu}
\affiliation{Department of Physics, University of Miami, Coral
Gables, Florida 33124}
\author{J. J. Neumeier}
\affiliation{Department of Physics, Montana State University,
Bozeman, Montana 59717}
\author{J.~L.~Cohn}
\affiliation{Department of Physics, University of Miami, Coral
Gables, Florida 33124}

\begin{abstract}
The dc magnetization ($M$) and electrical resistivity ($\rho$) as functions of
magnetic field and temperature are reported for a series of lightly electron doped
Ca$_{1-x}$La$_x$MnO$_3$($0\leq x\leq 0.10$) specimens for which
magnetization [Phys. Rev. B {\bf 61}, 14319 (2000)] and scattering
studies [Phys. Rev. B {\bf 68}, 134440 (2003)] indicate an
inhomogeneous magnetic ground state composed of ferromagnetic (FM)
droplets embedded in a G-type antiferromagnetic matrix.  A change in the
magnetic behavior near $x=0.02$ has been suggested to be the signature of
a crossover to a long-ranged spin-canted phase.
The data reported here provide further detail about this crossover in the magnetization, and
additional insight into the origin of this phenomenon through its manifestation
in the magnetotransport. In the paramagnetic phase ($T\geq 125$~K)
we find a magnetoresistance $\Delta\rho/\rho=-C(M/M_S)^2$ ($M_S$
is the low-T saturation magnetization), as observed in many
manganites in the ferromagnetic (FM), colossal magnetoresistance
(CMR) region of the phase diagram, but with a value of $C$ that is
two orders of magnitude smaller than observed for CMR materials.
The doping behavior $C(x)$ follows that of $M_S(x)$, indicating
that electronic inhomogeneity associated with FM fluctuations
occurs well above the magnetic ordering transition.
\end{abstract}

 \maketitle

\section{\label{sec:Intro} INTRODUCTION}

Inhomogeneous magnetic ground states, consisting of ferromagnetic (FM) or spin-canted clusters embedded
within an antiferromagnetic (AF) background, characterize both the lightly hole-doped\cite{Hennion,Terashita} (Mn$^{3+}$-rich)
and lightly electron-doped\cite{Chiba,Troyanchuk,Maignan,NeumeierCohn,Martin,Savosta,Mahendiran,Santhosh,GranadoRaman,Respaud,Aliaga,CohnNeumeier,LingGranado}
 (Mn$^{4+}$-rich) perovskite manganites.  There is growing consensus
that such inhomogeneity is intrinsic,\cite{Dagotto} an example of phase separation induced by a competition
between double-exchange (DE) and superexchange (SE) interactions between magnetic ions.

\begin{figure}
\includegraphics[width = 3.8in,clip]{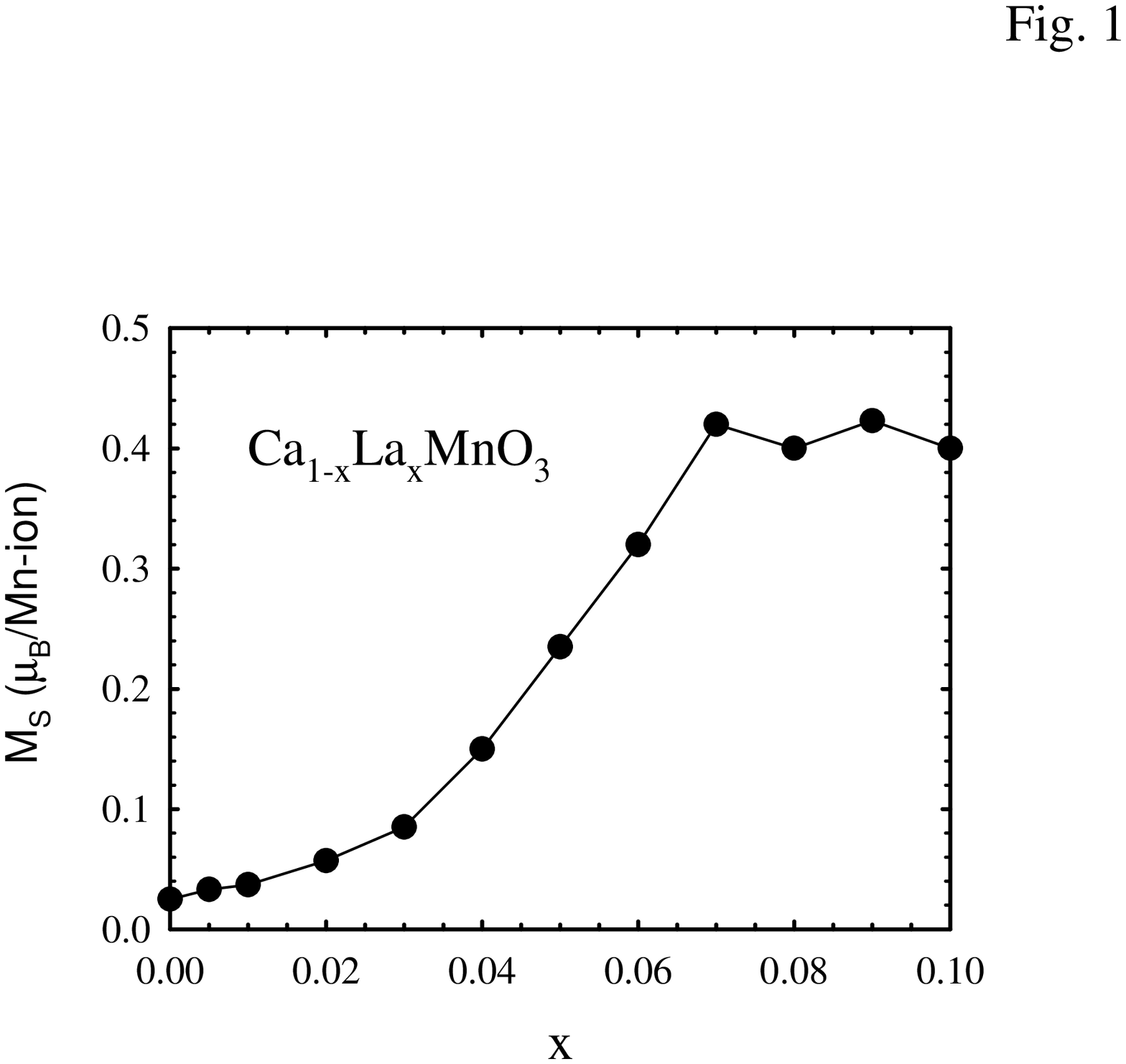}%
\caption{$T=5$K saturation magnetization {\it vs} $x$ from
Ref.~\protect\onlinecite{NeumeierCohn}.}
\label{MSvsX}
\vglue -.15in
\end{figure}

The anomalous magnetic behavior of the electron-doped compounds is
reflected in the appearance of a small FM magnetization that turns on below the antiferromagnetic N\'eel
temperature ($T_N$), saturates in an applied magnetic field of $H\geq 1-2$~T, and has a low-$T$ value ($M_S$)
that exhibits a change in slope as $x$ increases,\cite{NeumeierCohn,Mahendiran,Respaud,Aliaga} occurring near $x=0.02$ for
Ca$_{1-x}$La$_x$MnO$_3$ (Fig.~\ref{MSvsX}).
The origin of this slope change and its manifestation in magnetotransport is the focus of the present investigation.

A competition between DE and SE
underlies the magnetic phase behavior in electron doped
manganites. Recent neutron scattering studies\cite{LingGranado} of lightly doped Ca$_{1-x}$La$_x$MnO$_3$
compounds evidence a novel ground state wherein electron doping introduces FM polarons of nanometric
size that increase in density (but not size) with increasing $x$.  For $x\geq 0.06$ these studies indicate
the development of a long-range spin-canted state.  It was postulated that the crossover near $x=0.02$
reflects the large-scale aggregation of intermediate-sized spin-canted regions that coexist with isolated
FM droplets of smaller size.  Whereas magnetization measurements effectively sum over
all the magnetic species present, we anticipate that the magnetotransport will be more sensitive to the
hypothesized larger-scale spin-canted regions.

Here we examine measurements of magnetization and
magnetoresistance (MR) of Ca$_{1-x}$La$_x$MnO$_3$($0\leq x\leq
0.10$) in the range $1.5 {\rm K}\leq T\leq 250$~K and fields
$H\leq 9$~T.  The low-$T$ magnetization ($M$)curves, measured to
higher fields than previously reported,\cite{NeumeierCohn} reveal
a doping dependent susceptibility that can be deconvolved into
field-independent AF and field-dependent FM components, the latter
presumably reflecting the growth of FM and/or spin-canted domains.
The magnetotransport exhibits a crossover in behavior near
$x=0.02$ in both the magnetically ordered and PM phases. We
observe that $\Delta\rho/\rho=C(M/M_S)^2$ describes the PM phase
MR well, but with $C$ two orders of magnitude smaller than found
for CMR compounds due to the small volume fraction of DE bonds in
the present materials. Particularly interesting is the doping
behavior of $C$ which follows that of the low-$T$ $M_S$,
indicating that paramagnetic phase FM fluctuations have the same
inhomogeneous character as the magnetic ground state.

\section{\label{sec:Expt} EXPERIMENT}

Ca$_{1-x}$La$_x$MnO$_3$ polycrystals were prepared by standard
solid-state reaction; the preparation methods along with
magnetization and resistivity measurements are reported
elsewhere.\cite{NeumeierCohn} Iodometric titration indicated the
oxygen content of all specimens fell within the range 3.00$\pm
0.01$.  6-probe Hall-bar specimens of approximate dimensions $3\times 1\times 0.15\ {\rm mm}^3$
were prepared with silver paint contacts for dc Hall and magnetoresistivity (MR) measurements
in a 9T magnet. The magnetic field was applied perpendicular to the plane of the plate-like specimens
in which the current flowed.  Both current and field reversal were employed in the Hall measurements; the MR
was measured for both forward and reverse field orientations.
The temperature was controlled with a
Cernox sensor.  Magnetization and zero-field resistivity for these compounds have been reported
previously,\cite{NeumeierCohn} as well as thermopower and Hall mobility in the
paramagnetic phase.\cite{CohnPolarons} The latter measurements
indicate an electron density near room temperature in good agreement with values of $x$.
The very high resistivity
of the $x=0$ compound restricted measurements of its magnetotransport to $T\geq 75$~K and thus
it is excluded from the subsequent presentation.

\section{\label{sec:Results} RESULTS AND ANALYSIS}

\subsection{\label{AF-FM} Magnetically Ordered Phase ($T<T_N$)}

\subsubsection{\label{Chi} Magnetization }

\begin{figure}
\includegraphics[width = 3.88in,clip]{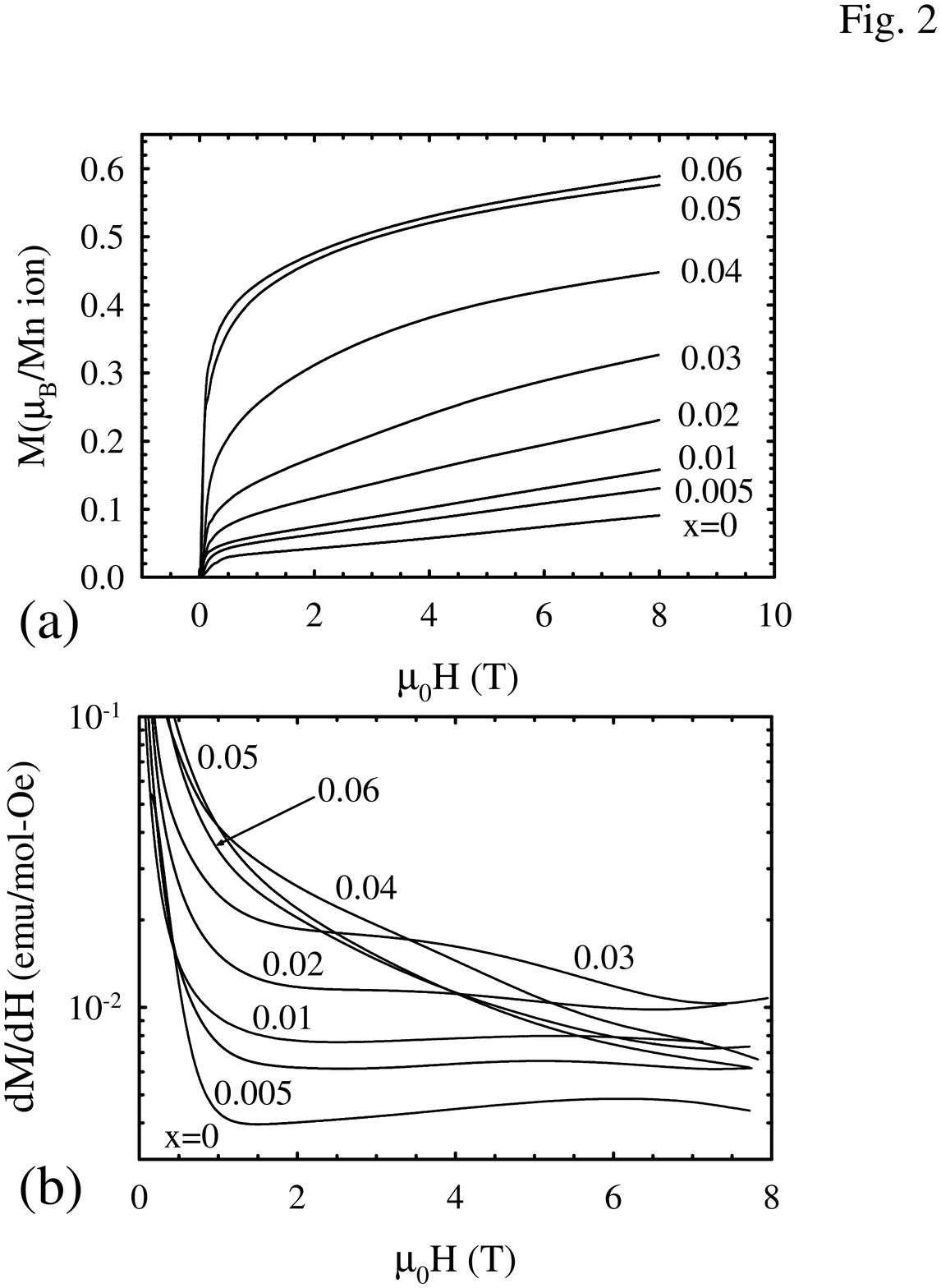}%
\caption{(a) Magnetization {\it vs} applied magnetic field for
Ca$_{1-x}$La$_x$MnO$_3$ polycrystals.  (b) Differential
susceptibility {\it vs} applied magnetic field determined from
data in (a).} \label{Magnetization}
\end{figure}
\begin{figure}
\includegraphics[width = 3.8in,clip]{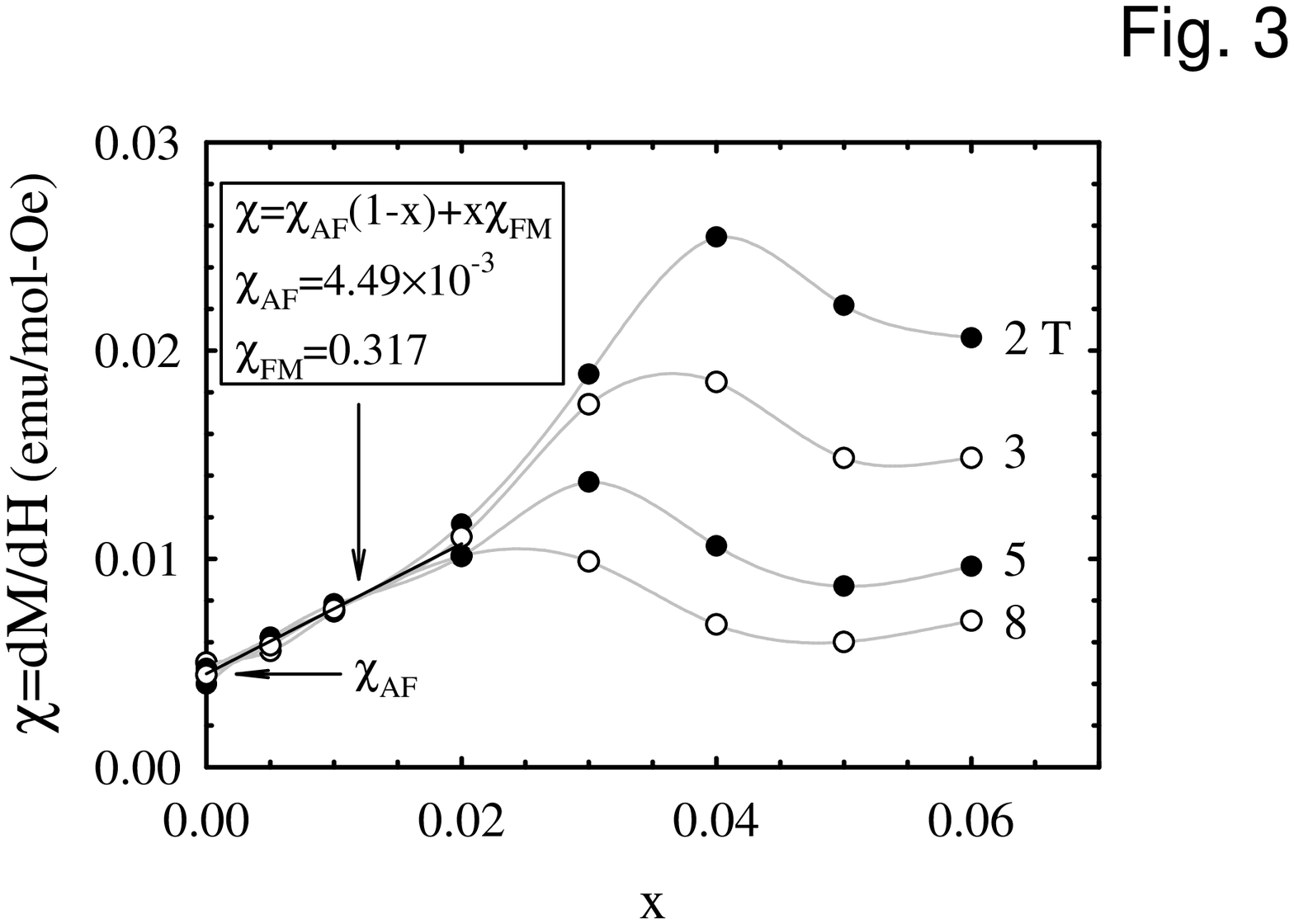}%
\caption{Magnetic susceptibility {\it vs} doping, at several values of magnetic field.
The solid line is a least-squares fit to the data for $x\leq 0.02$.
}
\label{ChivsX}
\vglue -.2in
\end{figure}

Figure~\ref{Magnetization}~(a) shows the $T=5$~K magnetization,
$M(H)$, for a series of specimens measured to higher field than
that reported previously in Ref.~\onlinecite{NeumeierCohn} and with a
data density in magnetic field that allows for a careful examination of
the differential susceptibility, $dM/dH$, shown as a function
applied field in Fig.~\ref{Magnetization}~(b).
These curves show an interesting trend: above $H\simeq 2$~T (a saturation field for
reorienting FM domains), $dM/dH$ is nearly
independent of field, increases approximately linearly with
doping up to $x=0.02$, and becomes strongly field dependent for $x\geq 0.03$.
The intermediate compositions, $x=0.02,\ 0.03$
exhibit $dM/dH$ curves that appear transitional between these behaviors: field-independent for $2\leq \mu_0H\leq 4$~T
and more strongly field dependent for $H\gtrsim$~4~T.  That a crossover in behavior occurs near
$x=0.02$ is more clearly seen in Fig.~\ref{ChivsX} where $dM/dH(x)$ is plotted for several fixed
values of magnetic field.

If contributions to the FM magnetization come from isolated
regions (e.g. FM droplets or spin-canted clusters) embedded in an
AF background, the doping-dependent susceptibility should be a
simple sum of terms weighted by their volume
fractions,\cite{AliagaChi}
\begin{eqnarray}
\chi=(1-x)\chi_{AF}+x\chi_{FM}
\end{eqnarray}

\noindent
where $\chi_{AF}$ represents the constant susceptibility of the AF background and
$\chi_{FM}$, in general field dependent, describes the growth in the isolated FM regions with applied field.
This simple model describes
the $dM/dH$ data at low doping ($x\leq 0.02$) quite well [solid line in Fig.~\ref{ChivsX}],
with $\chi_{AF}=4.49\times10^{-3}$~emu/mol-Oe and $\chi_{FM}=0.317$~emu/mol-Oe, both independent of field.
This value of $\chi_{AF}$ is in excellent agreement with the value extrapolated using Curie-Weiss fit
parameters established for a similarly prepared CMO specimen from $M(T)$ data\cite{NeumeierGoodwin}
at $H=2$~kOe in the range $200$~K~$\leq T\leq 400$~K.  The value of $\chi_{FM}$ corresponds to
$\sim 0.57$~$\mu_B$/Mn ion-kOe, i.e. for every 20 kOe of applied field
the magnetization increases by the equivalent of a spin-polarized electron for each doped electron.
The field independence of $\chi_{FM}$ is consistent with a FM contribution that is fully saturated
at $\mu_0H\geq 2$T.  At higher doping ($x>0.02$) $\chi(x)=dM/dH$ becomes strongly field dependent, with substantially
smaller values at higher fields.  This implies a field dependent $\chi_{FM}$ for $x>0.02$,
reflecting the gradual approach to saturation of the FM contribution with increasing field.

\subsubsection{\label{MR} Magnetotransport}
\begin{figure}
\includegraphics[width = 3.35in,clip]{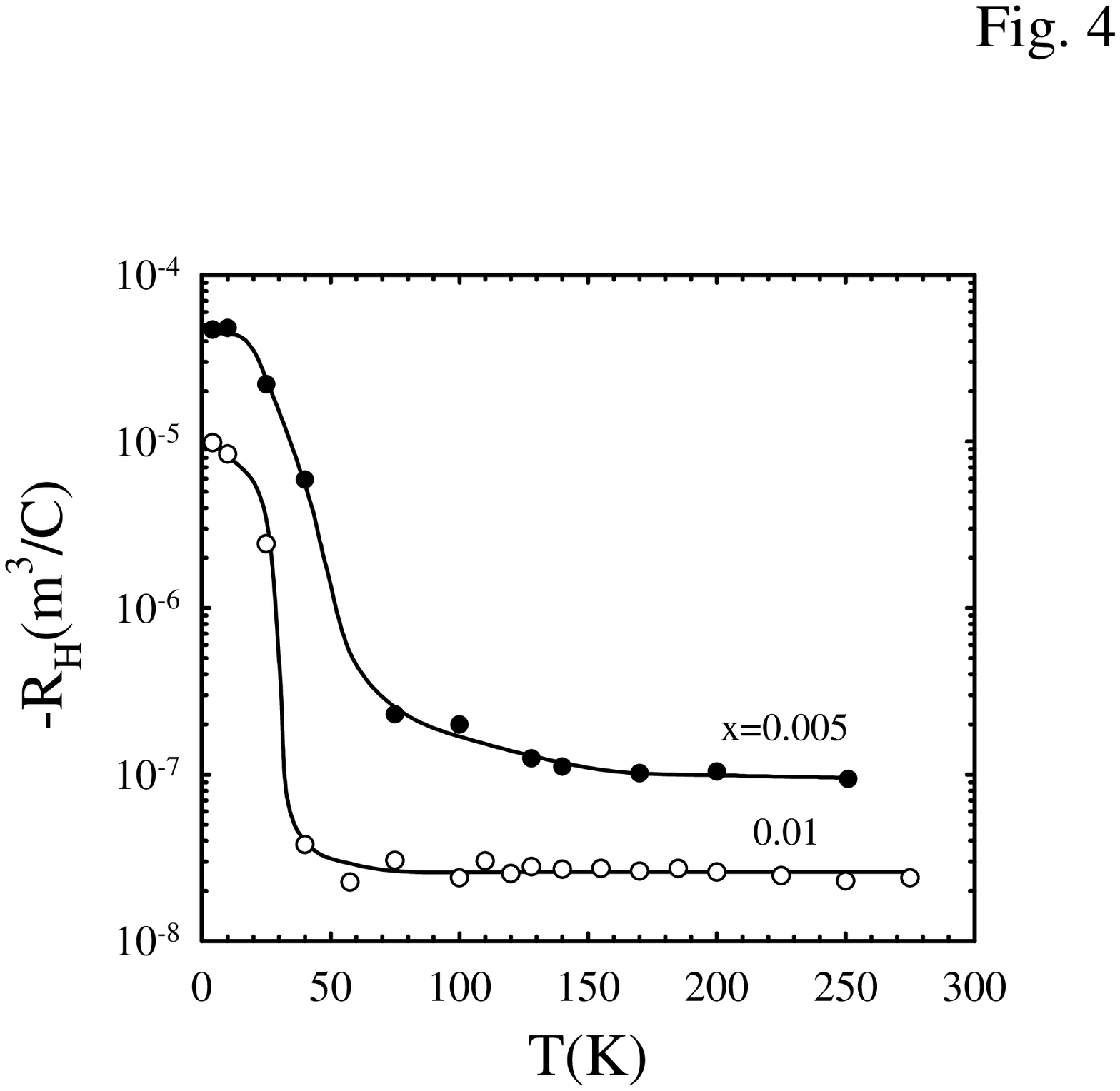}%
\caption{Hall number {\it vs} $T$ for $x=0.005,\ 0.01$ specimens.  Curves are guides to the eye.
}
\label{Hall}
\end{figure}

Prior transport measurements in the magnetic\cite{NeumeierCohn} and paramagnetic\cite{CohnPolarons} phases
establish the La-doped compounds as heavily-doped, lightly compensated $n$-type semiconductors
with modestly heavy (large-polaron) masses,
$m^{\ast}\sim 10m_0$.  The zero-field $\rho(T)$ for $T\lesssim 100$~K exhibits two
temperature regimes of simply activated behavior,  characterized by different
activation energies: \cite{NeumeierCohn} $E\sim 30$~meV and
approximately independent of $x$ ($40$~K$\leq T\leq 100$~K), and $\varepsilon\sim 0.1$~meV and dependent on the
saturation magnetization ($T\lesssim 10$~K).
The normal Hall effect for $x=0.005,\ 0.01$ could be reliably separated at low temperatures from
the anomalous Hall effect term because the latter saturated for $H\geq 2$~T like $dM/dH$ (Fig.~\ref{Magnetization}).
This was not the case for higher $x$, since a field dependent anomalous term throughout the available field range
is implied by the field-dependent $\chi$.  The Hall coefficient ($R_H$) is nearly constant for $T\geq 75$~K
for all specimens,\cite{CohnPolarons} and decreases for $x=0.005,\ 0.01$ by more than two orders of
magnitude at the lowest $T$ (Fig.~\ref{Hall}).  Thus the activation energy $E$ is attributable to a decrease in
mobility rather than carrier density.  The low-$T$ behavior of $R_H$ for $x=0.005,\ 0.01$ is consistent with a
freeze-out of electrons from the conduction band into La donor levels.  The small size of $\varepsilon$ and
observation that the thermopower tends toward zero\cite{NeumeierCohn,CohnPolarons} indicate that transport at the
lowest temperatures proceeds via carrier hopping within an ``impurity-band,'' likely comprised of La donor states.

\begin{figure}
\vglue -.1in
\includegraphics[width = 3.4in,clip]{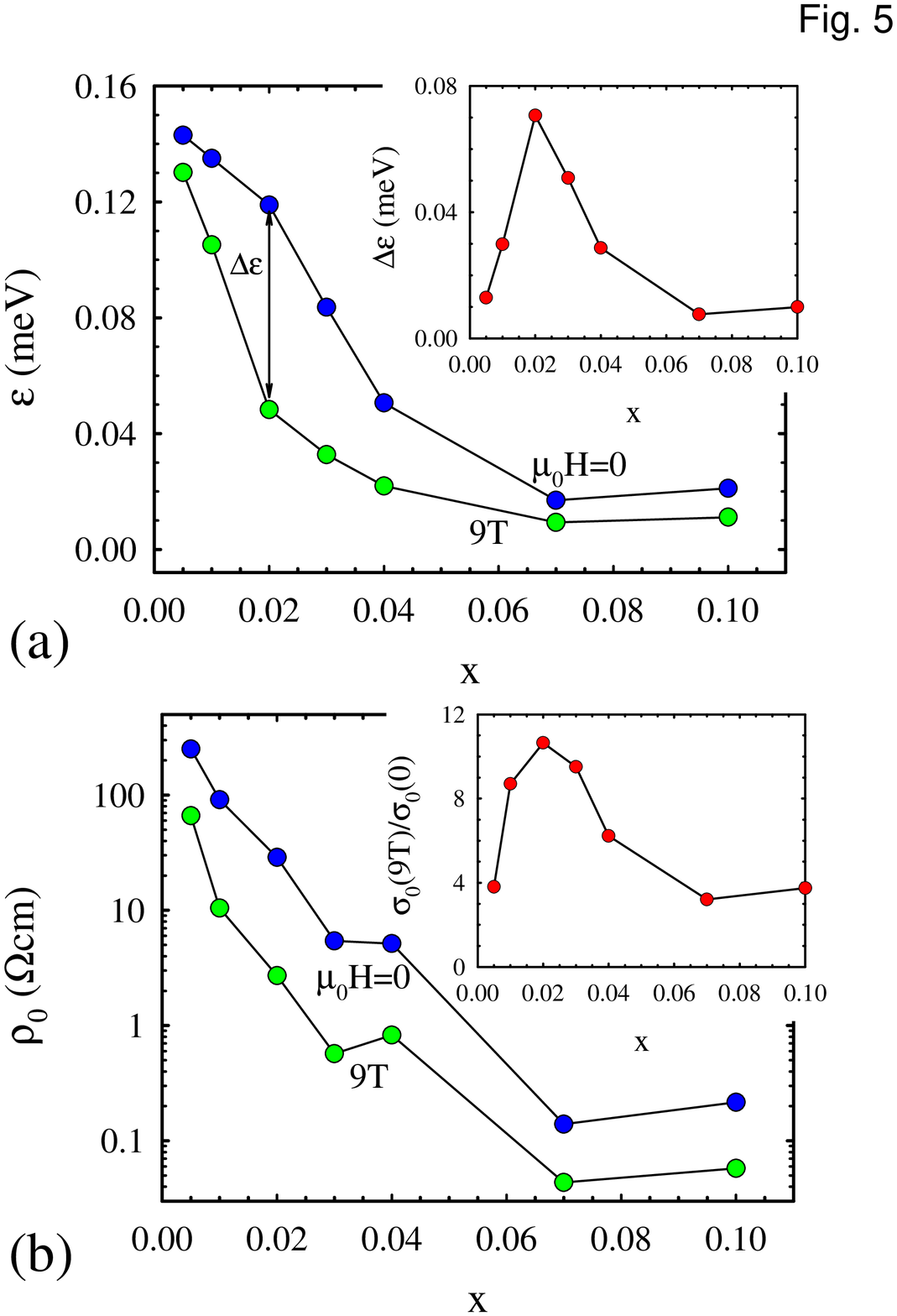}%
\caption{(a) Low-$T$ activation energy and (b) prefactor $\rho_0$ of the resistivity
{\it vs}~ $x$ for Ca$_{1-x}$La$_x$MnO$_3$ specimens at fixed fields $\mu_0H=0,\ 9T$.  Inset in
(a): $\Delta\varepsilon\equiv \varepsilon(0)-\varepsilon(9T)$, {\it vs} doping.
Inset in (b) $\rho_0(0)/\rho_0(9T)$ {\it vs x}.}
\label{ActEnergy}
\end{figure}

In spite of the correlation of
$\varepsilon$ with $M_S$ noted in Ref.~\onlinecite{NeumeierCohn},
values of $\varepsilon$ determined from $\rho(T)$ data at $H=2T$ differed little from zero-field values.
The 9-T fields employed for the present work
reveal a significant and systematic magnetic field dependence of both $\varepsilon$ and
the prefactor $\rho_0$ in $\rho=\rho_0\exp(\varepsilon/k_BT)$ [Fig.~\ref{ActEnergy}~(a) and (b), respectively]:  both
decrease substantially in a 9-T field, with maxima in $\Delta\varepsilon$ and $\sigma_0(9T)/\sigma_0(0)\equiv \rho_0(0)/\rho_0(9T)$
occurring at $x=0.02$ (insets, Fig.~\ref{ActEnergy}).
\begin{figure}
\includegraphics[width = 3.4in,clip]{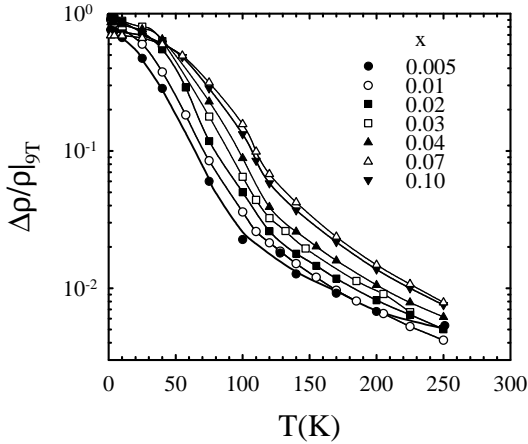}%
\caption{Normalized MR at $\mu_0H=9{\rm T}$ {\it vs} $T$ for Ca$_{1-x}$La$_x$MnO$_3$ specimens.
Curves are guides to the eye.}
\label{MaxMRvsT}
\end{figure}
\begin{figure}
\includegraphics[width = 3.8in,clip]{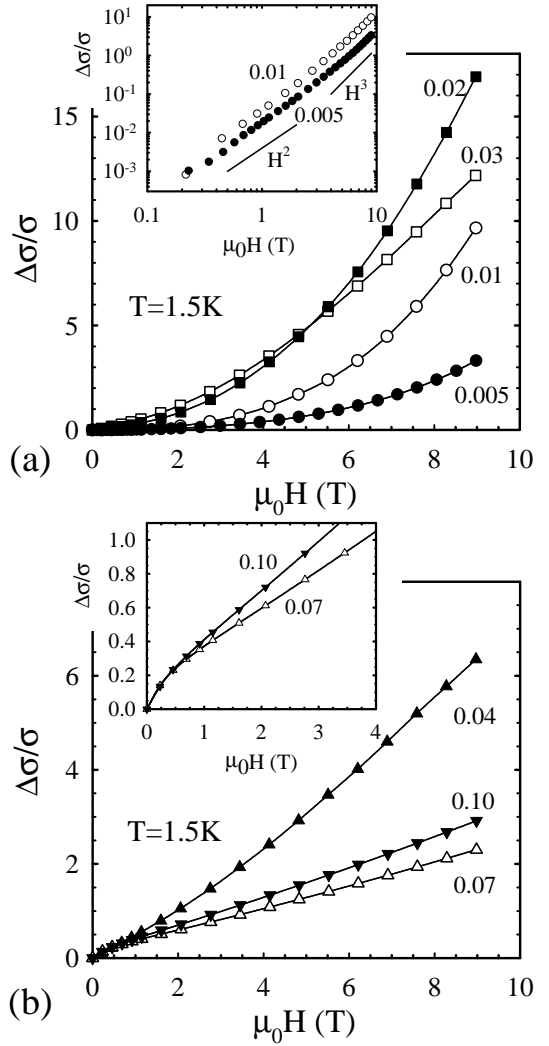}%
\caption{Normalized MC {\it vs} applied magnetic field at $T=1.5$~K for Ca$_{1-x}$La$_x$MnO$_3$
at low doping (a) $x\leq 0.03$ and higher doping (b) $x>0.03$.}
\label{lowTMC}
\end{figure}

\begin{figure}
\includegraphics[width = 3.6in,clip]{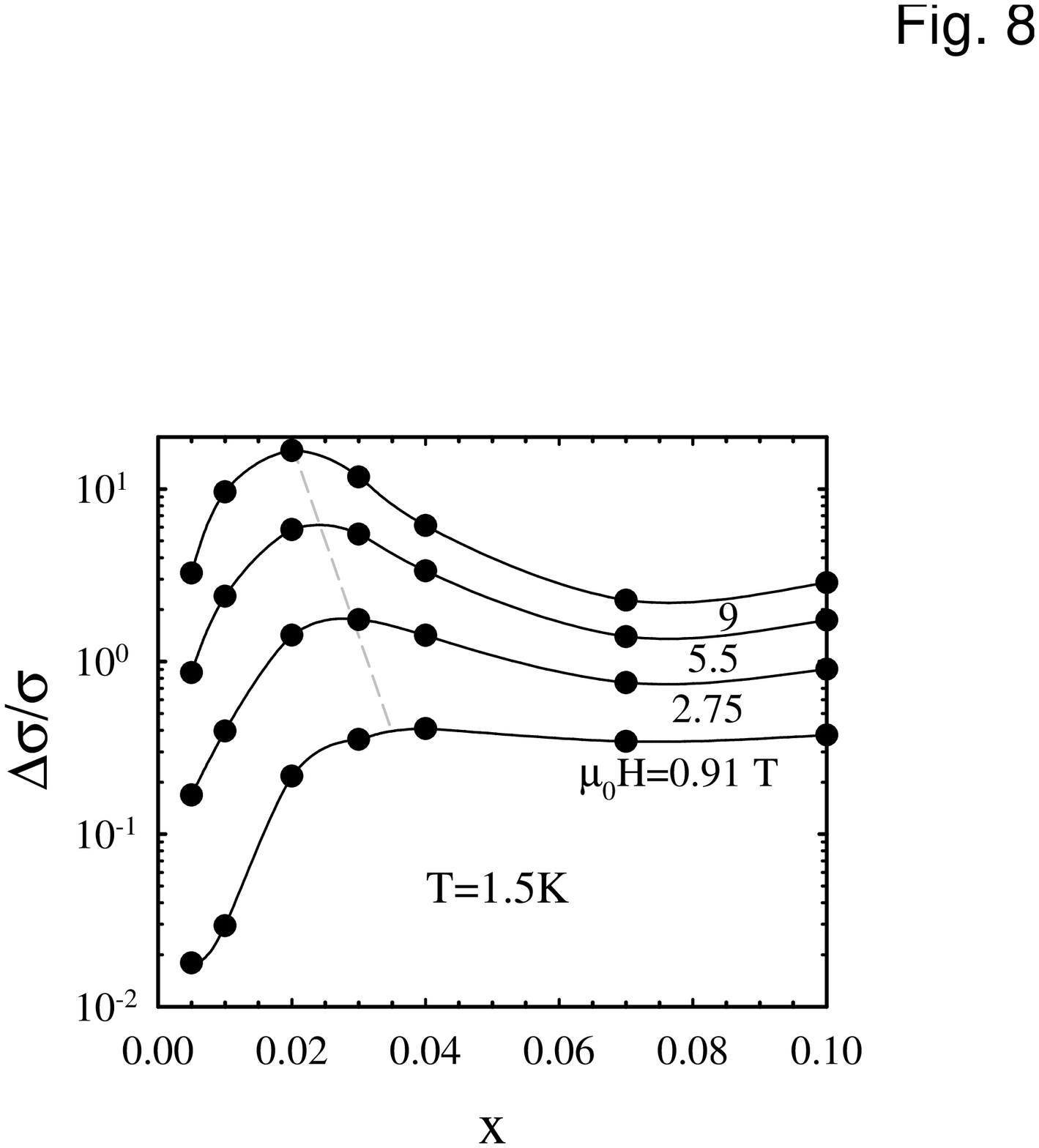}%
\caption{Normalized MC {\it vs x} at several values of applied magnetic field at $T=1.5$~K for Ca$_{1-x}$La$_x$MnO$_3$}
\label{lowTMCvsx}
\end{figure}

A negative MR, $\Delta\rho/\rho\equiv [\rho(H)-\rho(0)]/\rho(0)$, is observed for all specimens,
the magnitude of which increases with decreasing $T$ (Fig.~\ref{MaxMRvsT}), particularly at $T<T_N=110-125$~K.
The field dependence of the MR at the lowest $T$ also exhibits a crossover in behavior near $x=0.02$.
We choose to present the low-$T$ magnetotransport as magnetoconductivity (MC),
$\Delta\sigma/\sigma\equiv [\sigma(H)-\sigma(0)]/\sigma(0)$, since we find that it affords a more direct
comparison to the behavior of the susceptibility and the occurrence of a saturating FM component.
The $T=1.5$~K MC is shown in Fig.~\ref{lowTMC}.
For $x=0.005,\ 0.01$ [Fig.~\ref{lowTMC}~(a)] the MC is well represented by a power law, $\Delta\sigma/\sigma\propto H^m$
with $m\sim 2$ at low fields and $m\sim 3$ at the highest fields [inset, Fig.~\ref{lowTMC}~(a)].  There is no evidence for a
contribution that saturates like the FM component of magnetization (Fig.~\ref{Magnetization})
for $\mu_0H\gtrsim 2$~T, and thus the power-law behavior of the MC appears to be a characteristic of the lightly doped system.
For $x=0.07,\ 0.10$ the MC is larger at low field and smaller at high field [Fig.~\ref{lowTMC}~(b)] in comparison with
the lightly doped compounds, and
a contribution that saturates for H$\gtrsim 2$~T is evident [inset, Fig.~\ref{lowTMC}~(b)].  The MC's for compositions
$x=0.02,\ 0.03$ appear to be transitional in form between those of low and high doping.  The doping dependence of
the MC at fixed fields (Fig.~\ref{lowTMCvsx}) reveals a maximum magnitude for $x=0.02$ at the highest fields.

\subsection{\label{PM} Paramagnetic Phase ($T>T_N$)}

\begin{figure}
\includegraphics[width = 3.6in,clip]{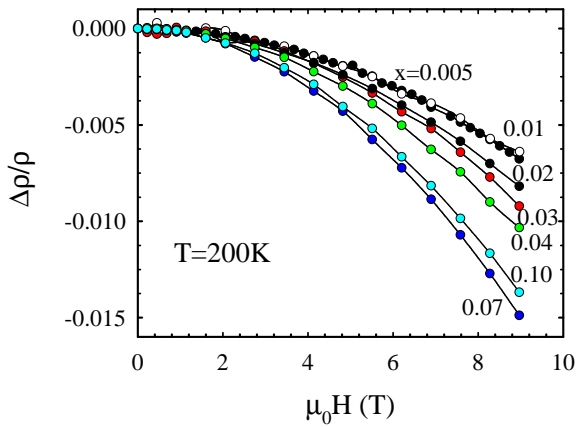}%
\caption{Normalized MR at $T=200$~K for Ca$_{1-x}$La$_x$MnO$_3$ specimens.}
\label{hiTMR}
\end{figure}

\begin{figure}
\includegraphics[width = 3.6in,clip]{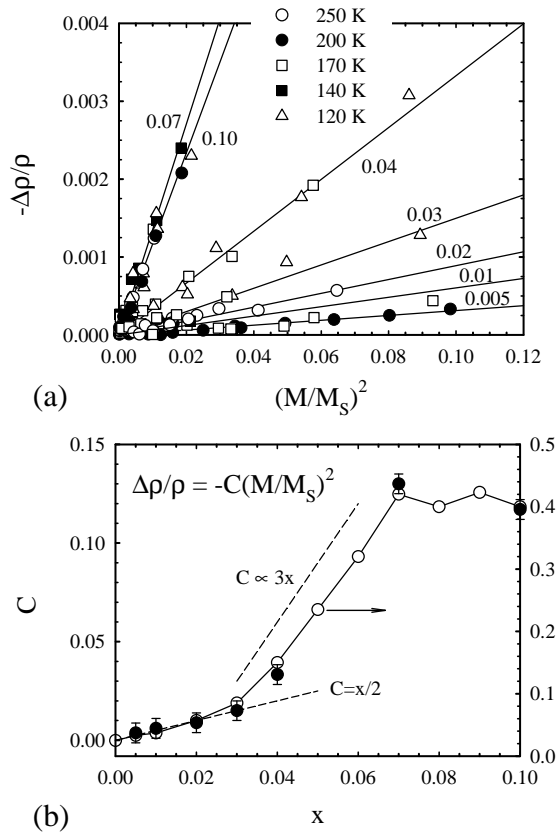}%
\caption{(a) Normalized MR vs $(M/M_s)^2$ in the PM phase for Ca$_{1-x}$La$_x$MnO$_3$ specimens at five
temperatures, and (b) Coefficient $C$ [slopes of data in (a); see text] (solid circles) and
$M_S$ (open circles and right ordinate) {\it vs x} for Ca$_{1-x}$La$_x$MnO$_3$ specimens.}
\label{CvsX}
\end{figure}

The MR in the PM phase is quadratic in field as shown for $T=200$~K in Fig.~\ref{hiTMR}.  Note that the
magnitude of the MR for $x=0.10$ is smaller than that for $x=0.07$, suggesting that the
MR is a simple function of the saturation magnetization, $M_S$ (Fig.~\ref{MSvsX}).  This motivates an
analysis along the lines
typical for CMR compositions where $\Delta\rho/\rho=-C[M(H,T)/M_S]^2$ is found to describe
combined magnetization and MR data near the Curie temperature.\cite{MRvsMsquared}
Figure~\ref{CvsX}~(a) shows that this scaling provides a good description of the MR data
at five temperatures in the PM phase.  Particularly interesting is the observation that
the doping variation of the slopes of these curves, $C(x)$, follows that of $M_S(x)$ remarkably
well [Figure~\ref{CvsX}~(b)], and thus also exhibits a slope change near $x=0.02-0.03$.
We find that $C(x)\simeq x/2$ for $x\leq 0.02$, and $C(x)\propto 3x$ for $0.02\leq x\leq 0.07$ (dashed lines).

\section{\label{sec:Disc} DISCUSSION}

The small angle neutron scattering (SANS) study by Granado {\it et al.}\cite{LingGranado} evidences the
presence of FM droplets of diameter 10~$\rm{\AA}$ that are decoupled from the G-type AF background.
Yet these droplets could account for less than half of the measured dc saturation magnetization at $x=0.02$.
Indeed, the density of droplets is about 60 times smaller than the electron
density (their minimum spacing at $x=0.02$ is $\sim 40\rm{\AA}$), implying that most electrons
reside outside of the FM droplets.  The development of a long-ranged
spin-canted state, evidenced in neutron scattering by Ling {\it et al.}\cite{LingGranado} for $x\geq 0.06$,
motivated these authors to postulate that the long-range spin-canted state emerges through a gradual
coalescence, with increasing $x$, of intermediate-scale spin-canted clusters not identified in the SANS
studies.  Thus it is proposed that most doped electrons reside in these spin-canted
clusters.

The changes in the doping and magnetic field dependencies of the susceptibility that occur
near $x=0.02$ (Fig.~\ref{ChivsX}) are consistent with a either a growth of the spin-canted cluster size or
the canting angle.  The mean distance between dopants,
$d=(4\pi n/3)^{-1/3}$ (where $n\simeq x/V_{f.u.}$
is the carrier density),\cite{CohnPolarons} is $\sim 8.4\ {\rm\AA}$ at $x=0.02$, about twice the Mn ion separation.
This is just the expected point of overlap for symmetric, 7-site FM polarons (a central Mn$^{3+}$ ion spin
aligned with those of its nearest neighbor Mn$^{4+}$), predicted to be the stable magnetic-polaron state
for this system.\cite{ChenAllen,Meskine}  But this appears to be a coincidence since,
as noted previously,\cite{NeumeierCohn} the low-doping regime ($x\leq 0.02$)
has $dM_S/dx\simeq 1\mu_B/$Mn-ion per doped electron
(Fig.~\ref{MSvsX}), much smaller than would be expected if each electron created an
isolated 7-site FM polaron.  The slope for $0.02\leq x\leq 0.07$ is close to that expected for isolated
7-site FM polarons, but in this regime long-ranged spin canting emerges.  Thus the correct picture is evidently
more complicated.

The transport data give further insight into the issue of magnetic inhomogeneity.
At the lowest $T$ the zero-field electrical conductivity for $x=0.005$ is higher than that for
$x=0$ by five orders of magnitude.\cite{NeumeierCohn}  This fact, along with the large, positive MC argue
against the coexistence, for $x\leq 0.02$, of the FM droplets with an unaltered G-type AF spin background;
isolated droplets separated by $\sim 40{\rm \AA}$ or more cannot dramatically improve electron transfer
relative to that of the background lattice.  Intervening non-percolating, spin-canted regions between
droplets appear to be the simplest modification capable of reconciling the magnetization, neutron,
and transport data.  The particular spin structure remains to be determined, but a plausible symmetric candidate
has a FM droplet surrounded by a spin-canted region.
This arrangement might naturally account for liquid-like correlations in the distribution of the droplets
observed in neutron scattering.\cite{LingGranado}

For impurity band conduction, $\rho_0$ and $\varepsilon$ represent the overlap of impurity-level wave functions and
their average energy difference, respectively.
The results of Fig.~\ref{ActEnergy} indicate that in the low-$T$ ordered state both of these are substantially
reduced in an applied field.  Since $\Delta\sigma/\sigma=[\rho_0(0)/\rho_0(H)]\exp(\Delta\varepsilon/k_BT)-1$,
the factor $\rho_0(0)/\rho_0(H)$ clearly predominates.  Thus increased
overlap of localized states underlies most of the MC and this effect is maximal
near $x=0.02$.

Regarding the PM phase MR, the empirical expression $\Delta\rho/\rho=-C[M(H,T)/M_S]^2$
has been widely employed for the FM CMR manganites for $M/M_S\lesssim 0.3$. Originally,
the more general expression, $\rho=\rho_0\exp{1-(M/M_S)^2}$, was proposed
to describe Eu chalcogenides,\cite{vonMolnar} with an activation energy reduced in
proportion to $M^2$.\cite{Bebenin}  Models based on DE\cite{Furukawa,InoueMaekawa} and
those invoking magnetization-dependent
variable range hopping barriers\cite{Coey,Wagner,Viret} have been proposed.
For metallic ferromagnets and magnetic semiconductors
this empirical correlation between MR and magnetization
emerges in generic models of magnetic scattering due to FM fluctuations.\cite{MajumdarLittleWood}
As noted above and in Ref.~\protect\onlinecite{CohnPolarons}, the
PM phases of the present compounds are best described as heavily-doped semiconductors with nearly
constant carrier densities and temperature-dependent mobilities.  Thus a magnetization-dependent
mobility would appear to be a more suitable description of the present system, with
FM fluctuations in the PM phase tending to reduce $\rho$ by providing DE pathways.
Regardless of whether a hopping barrier or scattering picture is more appropriate, it is interesting to
examine the magnitude and doping dependence of the coefficient $C$ for the electron-doped compounds
in comparison to the CMR manganites for which much data is available.

The magnitude of $C$ is quite small (CMR compounds have $C\sim 1-7$).\cite{MRvsMsquared}
This is to be expected given that for doping
$x$, the fraction of Mn$^{3+}$O$^{2-}$Mn$^{4+}$
bonds for which FM DE can operate is only $2x$, and thus Mn$^{4+}$O$^{2-}$Mn$^{4+}$ SE interactions
dominate the magnetic fluctuations.  Assuming that only the fluctuation contribution to the conduction
is field dependent, we have
$\Delta\rho/\rho\simeq 2x(\Delta\rho{MF}/\rho_{MF})$, where $\Delta\rho{MF}/\rho_{MF}$ is the hypothetical
MR that would occur in a uniform system where all bonds are DE active.  Thus the value $C\approx 0.13$ found
at $x=0.07$ is roughly equivalent to $C\sim 0.13/2x\simeq 1$ for a uniform system, comparable to values for
pure DE systems.

The finding $C(x)\propto M_S(x)$ (Fig.~\ref{CvsX}) provides compelling evidence that the PM-phase MR is
associated with FM fluctuations.  Furthermore, since the crossover behavior near $x=0.02-0.03$ is also evident
in $C(x)$, we conclude that these fluctuations have the same inhomogeneous structure present in the ground state:
fluctuating FM droplets and/or spin-canted clusters begin developing well above the ordering temperature.
It is reasonable to view this as a natural consequence of the need to maintain local charge neutrality;
doped electrons tend to spend more time near La dopants with decreasing temperature, and eventually
become trapped in the vicinity of these donor sites at low $T$.

In summary, magnetization and magnetotransport in electron-doped Ca$_{1-x}$La$_x$MnO$_3$
($0\leq x\leq 0.10$) evidence a crossover in behavior near $x=0.02$ in both magnetically-ordered and PM phases.
These phenomena are attributed to an inhomogeneous magnetic structure of both the ground state magnetism
and the FM fluctuations at $T>T_N$.  The present data lend support to the scenario motivated by recent neutron
scattering studies,\cite{LingGranado} that non-percolating, spin-canted
clusters coexist with smaller-scale FM droplets in this system.

\section{ACKNOWLEDGMENTS}

This material is based upon work supported by the National Science Foundation under grants
DMR-0072276 (Univ. Miami) and DMR-9982834 (Montana State Univ.).

%

%

%

\end{document}